\documentclass[12pt]{iopart}

\usepackage{graphicx}
\usepackage{color}

\newcommand{\muinit}{\mu_{\mathrm{init}}}
\newcommand{\muc}{\mu_{\mathrm{C}}}
\newcommand{\ecut}{\epsilon_{\mathrm{cut}}}

\begin{document}
\paper[Growth of a Bose-Einstein condensate: Comparison of theory and 
experiment]
{Growth of a Bose-Einstein condensate:\\
A detailed comparison of theory and experiment}
\author{M J Davis\dag\ddag \ and C W Gardiner\S}
\address{\dag\ Clarendon Laboratory, Department of Physics, University of 
Oxford,
Parks Rd, Oxford, OX1 3PU, United Kingdom}
\address{\ddag\ Department of Physics, University of Queensland, St Lucia, QLD 
4072,
Australia}
\address{\S\ School of Chemical and Physical Sciences,
Victoria University of Wellington, Wellington, New Zealand}
\ead{m.davis2@physics.ox.ac.uk}

\begin{abstract}

We extend the earlier model of condensate growth of Davis \etal [Davis MJ,
Gardiner CW and Ballagh RJ 2000 \PR A \textbf{62} 063608] to include the
effect of gravity in a magnetic trap.  We carry out calculations to model the
experiment reported by K\"{o}hl \etal [K\"{o}hl M, Davis MJ, Gardiner CW,
H\"{a}nsch T and Esslinger T, \textit{Preprint} cond-mat/0106642] who study the
formation of a rubidium Bose-Einstein condensate for a range of evaporative
cooling parameters.  We find that in the regime where our model is valid, the
theoretical curves agree with all the experimental data with no fitting
parameters.  However, for the slowest cooling of the gas the theoretical curve
deviates significantly from the experimental curves.  It is possible that this
discrepancy may be related to the  formation of a quasicondensate.
\end{abstract} 

\section{Introduction}

The process by which a Bose-Einstein condensate (BEC) forms from a
non-equilibrium thermal vapour is an important question in finite temperature
field theory.  Before the first observations of condensates \cite{JILA, MIT,
RICE}, estimates for the characteristic time of formation varied dramatically
(eg see  discussion in \cite{stoof_long}).  However, more recently  the quantum
kinetic theory of Gardiner and Zoller \cite{QKT}  has resulted in quantitative
predictions that can be compared with experimental data.

Until recently  the only experimental study of the process of condensate
formation  was that performed by Miesner \etal \cite{MIT_growth} in the group
of Ketterle at MIT.  In these experiments a cloud of sodium atoms was cooled to
just above the BEC transition temperature, before the high-energy tail of the
distribution was quickly removed by a rapid sweep of the rf field frequency.
The resulting dynamics lead to the formation of a condensate, with the
observation of characteristic S-shaped growth curves.

Before the first measurements of condensate growth, a quantitative  prediction
of the growth rate was presented in \cite{BosGro1}, in which a condensate was
assumed to form from contact with a thermal bath below the transition
temperature. In order to give a simple estimate of the rate constant, the
approach was greatly simplified and had several limitations---the most
important being the use of a Maxwell-Boltzmann rather than Bose-Einstein
distribution function for the thermal bath. However, it gave a good qualitative
prediction of the  general shape and order of magnitude of the growth rate
later observed.

This model was soon extended to include both Bose-Einstein  statistics for the
thermal bath of atoms,  and the dynamics of the lowest lying quasiparticle
levels above the condensate \cite{BosGro2,QKVI}.  The picture  was of a
condensate band in contact with a ``supersaturated'' thermal cloud, and the
assumptions of the model matched the experimental conditions realized at MIT
quite closely.  A comparison of the theoretical predictions with experimental
data was in good agreement at higher temperatures; however, at lower
temperatures there was some  discrepancy---the experimental growth rate
appeared to be about three times too  fast.

In further development, this model was again extended to include the dynamics
of the evaporative cooling in \cite{QKVII}.  While this predicted faster 
growth in some circumstances as compared to the simpler model, it did not occur
in  the parameter regime of the MIT experiment and thus the discrepancy
remained.  In addition, the necessary experimental data for a proper
theoretical treatment  was not available.  It was concluded that for a rigorous
comparison with theory it was necessary for further experiments to be carried
out with all relevant data recorded.  The same conclusion was reached in a
similar calculation by Bijlsma \etal \cite{Bijlsma}.

Recently a  carefully controlled study of condensate formation in a rubidium
vapour was carried out by K\"{o}hl \etal \cite{koehl} in Munich.  They used a
different cooling scheme  from that used in the the MIT experiment---instead 
of a rapid rf sweep after cooling to near the transition temperature, they
turned on a constant frequency rf field.  This allowed them to vary the rate of
evaporative cooling by  changing the rf frequency between experiments, and they
report their growth curves, initiation times, and growth rates  in
\cite{koehl}. Most importantly, the Munich group  measured all the relevant
theoretical parameters,  so that calculations with \emph{no free
parameters} can be carried out.  

This paper is organized as follows.  In section \ref{model} we summarize the
theoretical model of \cite{QKVII}, before describing the extensions necessary
to model the experiments of  K\"{o}hl \etal \cite{koehl} and discussing the
validity of the approximations made.  In section \ref{results}  we investigate
the effect of the model extensions  on condensate growth as compared to the
earlier calculations in \cite{QKVII}, and then compare the results of the 
model to the experimental data and discuss their implications.  Finally,  our
conclusions are presented in section \ref{conclusion}.

\section{\label{model} Theoretical model}

The model we use to simulate the experiments of K\"{o}hl \etal is 
described
fully in reference \cite{QKVII}.  The description is based on quantum kinetic
theory \cite{QKT}, but essentially  reduces to solving  a modified 
ergodic
quantum Boltzmann equation (MQBE) that assumes the
distribution function depends only on energy.  Our method makes the additional
assumptions that:
\begin{itemize}
\item[(i)] The condensate wave function and energy eigenvalue [the 
condensate chemical potential $\muc(n_0)$] are given by the 
solution of the time-independent Gross-Pitaevskii equation with $n_0$ atoms.
We assume
that the growth of the condensate is adiabatic and
that its shape is always well-described by the  Thomas-Fermi wave function.
\item[(ii)] The excited states above the condensate are the quasiparticle 
levels appropriate to the condensate wave function, leading to a density 
of states for the system that is substantially modified from the
non-interacting case. We use a 
particle-like density of states, thus neglecting any specifically 
quasiparticle behaviour, whose effects are expected to be minor \cite{dalfovo}.
\end{itemize}

To model the MIT experiments, the simulations were begun with an initial
distribution  truncated at an energy $\ecut$ and so the process of
atom loss during the evaporative cooling did not need to be included.  However,
as the Munich experiment  involves continuous evaporative cooling, it must be
included in the  simulation.  To do so we solve the effective MQBE
\begin{eqnarray}
g_n \frac{\partial f_n}{\partial t} &=& 
\frac{8 m a^2}{\pi \hbar^3}
\sum_{pqm} g_{\mathrm{min}} 
\delta(\epsilon_p + \epsilon_q -\epsilon_m - \epsilon_n)
\nonumber\\
&& \times\left\{
\beta_n f_p f_q (1+f_m)(1+f_n) - f_m f_n (1+f_q) (1+f_q)
\right\}
\label{eqn:MQBE}
\end{eqnarray}
where $a$ is the \emph{s}-wave scattering length, $m$ is the atomic mass,
$f_n \equiv f(\epsilon_n)$ is the distribution function, $g_n \equiv
g(\epsilon_n)$ is the density of states, and  $g_{\mathrm{min}}$ is the density of
states of the minimum energy particle participating in the  collision. The
quantity $\beta_n$ takes account of the evaporative cooling---it is one if the
energy $\epsilon_n \ge \ecut$, and zero otherwise.

There is, however, another effect that must be taken account of at low
temperatures.  The trapping potential that the atoms experience is due to not
only the applied magnetic field, but also the gravitational potential. While
gravity does not change the shape of the trapping potential, it shifts the
minimum of the trap away from the minimum of the magnetic field.  This has
important consequences for the evaporative cooling of the cloud.  Before we 
describe this further, however, we summarize the initial parameters of the
experiment we are modelling so we can quantitatively discuss the  magnitude of
the effect.

\subsection{\label{expt} Experimental summary}

The magnetic trap used by  K\"{o}hl \etal is well approximated by a 
cigar-shaped harmonic potential with trapping frequencies $\omega_x = \omega_z
= 2\pi \times 110$ Hz, $\omega_y = 14$ Hz, with a  geometric mean frequency of
$\bar{\omega} = (\omega_x \omega_y \omega_z)^{1/3} = 2\pi \times 55.3$ Hz. 
They begin their growth experiments  with a cloud of $N_i = 4.2 \pm 0.2 \times
10^6$  atoms of $^{87}$Rb trapped in the $|F=1,m_F = -1\rangle$ hyperfine
ground state, cooled to a temperature of $640 \pm 30$ nK giving an initial
chemical potential of $\muinit \approx -300 \hbar \bar{\omega}$. The rf fields
they applied to their cloud for which a condensate was observed to form were
between 2000--2090 kHz, corresponding to  
$0.92 \le \eta \equiv \ecut / k_B T_i \le 4.62$.

\subsection{Gravitational sag of the trapped atomic cloud}

\begin{figure}
\includegraphics{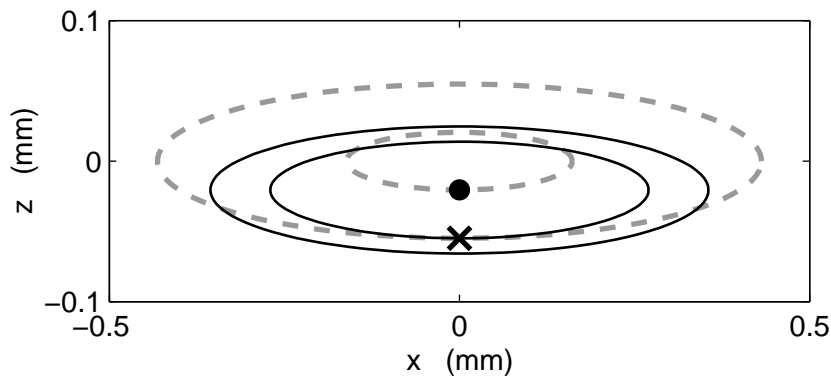}
\caption{\label{fig:sag} An indication of the sag of the  trapping potential as
compared to the magnetic field equipotentials.  The dashed gray lines indicate
the magnetic equipotentials corresponding to an rf field frequency of 1955 kHz
(inner) and 2090 kHz  (outer). The solid dot indicates the centre of the atomic
cloud.  The solid lines represent the spatial bounds of atoms with an energy
of  4.62$k_B T_i$ (inner) and  8$k_B T_i$ (outer).  The cross marks the
innermost intersection of the rf  field with atomic cloud equipotentials,
determining the quantity $\eta = 4.62$.   However, as can be seen there can be
atoms with energy 8$k_B T_i$ in orbits that will not  be ejected from the trap
via spin flips.}
\end{figure}

By including the  effect of gravity, we find that the trapping potential the
atoms experience is given by
\begin{equation}
V(\mathbf{r}) = \frac{m}{2}
\left[\omega_x^2 x^2 + \omega_y^2 y^2 + \omega_z^2 (z + A)^2 \right]
\end{equation}
where  the origin is defined as the minimum of the magnetic field, and $A = g 
/ \omega_z^2 = 20.5$ $\mu$m is the sag of the atom cloud below the origin. 
This situation is illustrated in figure~\ref{fig:sag}.

The sag of the atomic cloud has important consequences for evaporative cooling
at low temperatures.  In figure \ref{fig:sag} the solid dot represents the
centre of the atomic cloud, while the dashed grey curves represent magnetic
field equipotentials.  The innermost corresponds to an rf field frequency of
1955kHz, which was determined to be the minimum of the trap by atom laser
output coupling \cite{koehl}.  The outermost dashed grey curve represents an rf
field  frequency of 2090 kHz applied to the system, and all atoms that cross
this surface will be quickly ejected from the trap.

By considering the intersection of the equipotentials of the atomic cloud 
with the evaporative cooling surface, we find that all atoms with an energy
less  than $\ecut = 4.62 k_B T_i$ will remain trapped.  However,
not all atoms with higher energy will be ejected---as can be seen by
considering the  equipotential corresponding to $\epsilon =  8 k_B T_i$.  Atoms
with this energy in a horizontal orbit will not cross the evaporative cooling
surface, and hence  will remain trapped at least until they suffer another
collision.

\subsubsection{Inclusion of sag in model}

One of the limitations of our theoretical model is that the distribution
function of the gas is assumed to be ergodic---that is, all states of the gas
with the same energy are assumed to have the same occupation.  This obviously
cannot hold rigorously in this situation, where atoms of the same energy will
be ejected depending on orientation of their orbit.  However, the effect of 
the sag of the cloud can be included in the model, if not entirely
rigorously.  

We proceed to calculate the fraction of atoms of a given energy that will
remain trapped during the application of a fixed frequency rf field, and use
this as our function $\beta(\epsilon_n)$ in equation (\ref{eqn:MQBE}).  As the
hottest atoms are ejected from the trap it is reasonable to assume that they
can be treated as being non-interacting.  Indeed, we found in reference
\cite{QKVII} that the density of states we use  is not greatly  altered from
the non-interacting case for energies larger than about three times the
condensate eigenvalue, which in these calculations never exceeds $45 \hbar
\bar{\omega}$.  In comparison the minimum energy for ejection is about $k_B T_i
\approx 240 \hbar \bar{\omega}$ so this approximation  does seem reasonable.

While it is possible to write down an integral describing the total number of
states of a given energy in phase space that will remain trapped, it is not
possible to  to give an analytic expression for this  quantity. Instead we
proceed using a Monte Carlo simulation of  non-interacting particles in the
trap.  For each energy we populate the initial states at random and then follow
the trajectories in time, removing each  that crosses the evaporative cooling
surface.  After a sufficiently long period we determine the proportion that
remain trapped.

\begin{figure}
\includegraphics{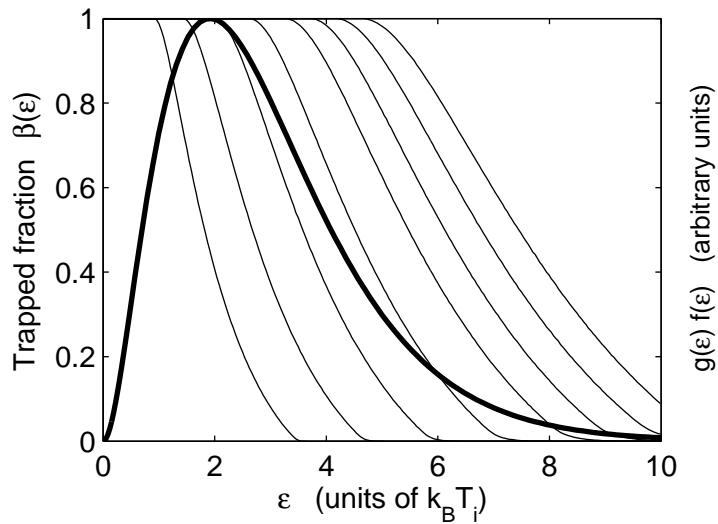}
\caption{\label{fig:loss2} Curves showing the fraction of trapped atoms for a
given energy with the application of rf fields: from left to right $\nu = $
2000, 2015, 2030, 2045, 2060, 2070, 2080, 2090 kHz.
The solid black line is a plot of the initial
distribution function of the gas $g(\epsilon) f(\epsilon)$ at $T_i = 640$ nK.}
\end{figure}

The curves we have calculated are illustrated in figure~\ref{fig:loss2} as a
function of the applied frequency of the rf field.  For comparison
we also plot the initial distribution function of the cloud. We can see
that for all applied fields the region where a finite fraction of atoms  is
trapped is quite wide, and therefore this effect is important for the
experiment we are considering here.  

\subsubsection{Validity of the model of the sag}

The inclusion of the function $\beta(\epsilon_n)$ in equation (\ref{eqn:MQBE})
relies on two approximations.  The first is that  an 
atom gaining an energy
higher than $\ecut$  is equally likely to enter any region of
phase space available to it.  This should be reasonable, as most collisions
that result in one particle entering this region will occur between two atoms
with energies less than $\ecut$, where the distribution function
should be ergodic.  The second assumption is that non-ergodicity of the levels
above $\ecut$ will not have a significant effect on the
calculation.  This remains unproven---however, it could be tested via Monte
Carlo simulations of evaporative cooling well above the BEC transition.

\subsection{Other effects}

A further measurement reported by K\"{o}hl \etal is a drift in the magnetic
field due to heating in the coils in the experiment, equivalent to a linear
decrease in the rf field frequency of 5 kHz s$^{-1}$.  This is easily included
in our model by making the function $\beta(\epsilon_n)$ time dependent.

Another factor that we include in our simulations is the loss of condensate
atoms due to three body processes, via the rate
\begin{equation}
\frac{d n_0}{dt} = - K_3 \int d^3 \mathbf{x} [n(\mathbf{x})]^3,
\end{equation}
where $n(\mathbf{x})$ is the condensate density, and $K_3$ is the three body
loss coefficient.
Using the Thomas-Fermi profile for the condensate density we find \cite{ENS}
\begin{equation}
\frac{d n_0}{dt} = - K_3 \frac{15^{4/5}}{168\pi^2} 
               \left(\frac{m \bar{\omega}}{\hbar \sqrt{a}}
	       \right)^{12/5} n_0^{9/5},
\end{equation}
where loss processes involving thermal cloud atoms have been neglected.  This
should be a reasonable approximation, as although such losses are
enhanced by a factor of 3!, the density of the thermal cloud is significantly
less than that of the condensate.  We use a value of 
$K_3 = 5.8 \times 10^{-30}$ cm$^6$ s$^{-1}$
 for the hyperfine state $|F=1,m_F = -1\rangle$ as reported by 
JILA \cite{JILA_3body}.

\section{\label{results} Results}

In this section we begin our simulations with the initial conditions as
reported by K\"{o}hl \etal and summarized in section \ref{expt}, and use a
scattering length for $^{87}$Rb of $a = 110 a_0$.   Note that this value
is subject to an uncertainty of a few percent and this could have a
effect on the results, mainly through a scaling of the time axis.

\subsection{\label{effect} Consequences of the trap sag and magnetic field 
drift}

\begin{figure}
\includegraphics{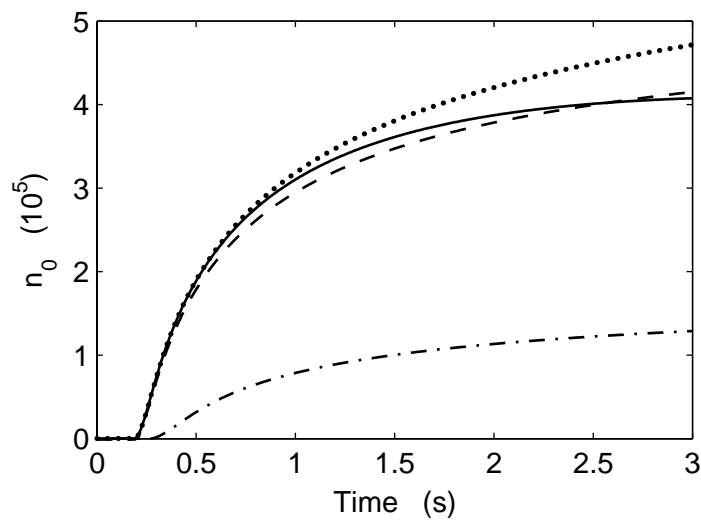}
\caption{\label{fig:sag_effect} Growth curves from an initial condition of 
$N_i = 4.2 \times 10^6, T_i = 640$ nK, $\muinit = -300 \hbar\bar{\omega}$, 
with an applied rf field of 2015kHz.  The dot-dash line indicates the predicted
condensate formation if all atoms above an energy of $\ecut /  (k_B
T_i) = 1.43$ are removed from the trap.  The dashed line takes into  account
the trap sag, and proportion of atoms lost at each energy is displayed in 
figure \protect\ref{fig:loss2}.  The dotted line also includes the
effect of  the drift of the magnetic field, and finally the solid line
additionally includes three body loss from the condensate. }
\end{figure}

The offset of the atomic trap from the mininum of the magnetic field has a
significant quantitative effect on the resulting growth curves, and this is
illustrated  in figure \ref{fig:sag_effect} for the initial condition  $N_i =
4.2 \times 10^6$, $T_i = 640$ nK and an applied rf field of 2015 kHz.   This
gives a minimum energy of atoms to be lost from the trap of 
$\ecut /  (k_B T_i) = 1.43$ at $t = 0$ s.

If \emph{all} atoms above the energy  $\ecut$ are continuously
removed  from the trap (dot-dash curve) [as if both the trap and magnetic field
equipotentials were  concentric], then the resulting condensate is much smaller
than is  predicted if we include the effects of the trap sag (dashed line). 
This is because such a heavy cut into the cloud removes a large proportion of
the initial number of atoms; however, the final condensate fractions for both
curves are similar.  The further inclusion of the magnetic field drift (dotted
line) has only the effect of making the final condensate slightly
larger---easily understood as this evaporatively cools the cloud further.
Finally, including three body loss from the condensate makes the growth curve
start to level off once $n_0 \approx 3 \times 10^5$.  The same qualitative
behaviour is observed for all values of the rf field.

If a similar experiment to those carried out by Miesner \etal \cite{MIT_growth}
at MIT was performed with this initial condition---all atoms above
$\ecut /  (k_B T_i) = 1.43$ are removed but with the rf field turned
off at $t=0$ s---then no condensate is observed to form.  This is because the
initial cloud is sufficiently far from the transition point  that this single
truncation cools the cloud from  $\muinit = -300 \hbar\bar{\omega}$ to $\muinit
\approx 0$, just before condensation occurs.

\subsection{\label{trends} Trends in the theoretical data}

\begin{figure}
\includegraphics{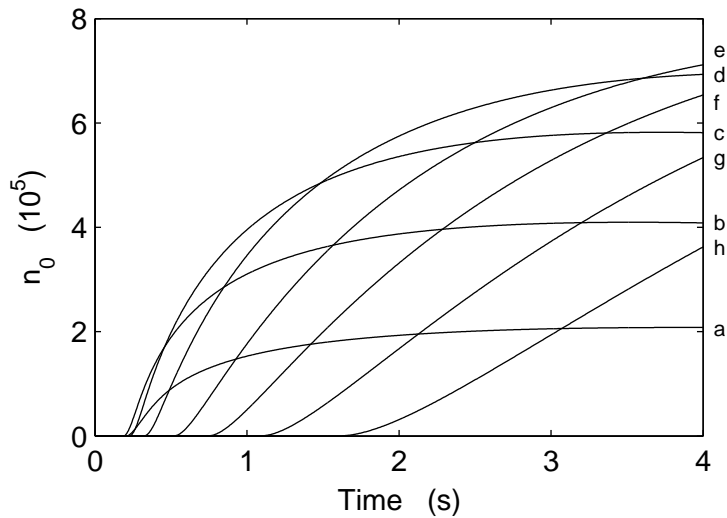}
\caption{\label{fig:T640N42} Growth curves from an initial condition of 
$N_i = 4.2 \times 10^6, T_i = 640$ nK, $\muinit = -300 \hbar\bar{\omega}$, for
rf fields of (a) 2000, (b) 2015, (c) 2030, (d) 2045, (e) 2060, (f) 2070,
(g)  2080, (h)  2090 kHz.}
\end{figure}

We now present the theoretical predictions for all experimental rf frequencies
in figure \ref{fig:T640N42}, including  the effects of all of: Trap sag,
magnetic field drift, and three body loss from the condensate.  We  show these
on the same figure for comparison of time scales---in the comparison with
experimental data below  we show only single curves on each graph.

These curves show the expected behaviour---the fastest evaporative  cooling
generally results in a shorter initiation time and more rapid initial growth. 
However, because the fastest evaporative cooling initially loses a large 
number of atoms without any collisions (the cloud is simply truncated, rather
than collisions causing atoms to be evaporated), this results in smaller 
condensates. This is the behaviour that was observed in the MIT-style
simulations performed in \cite{Bijlsma}.

\subsection{Comparison with experimental data}

\begin{figure}
\includegraphics{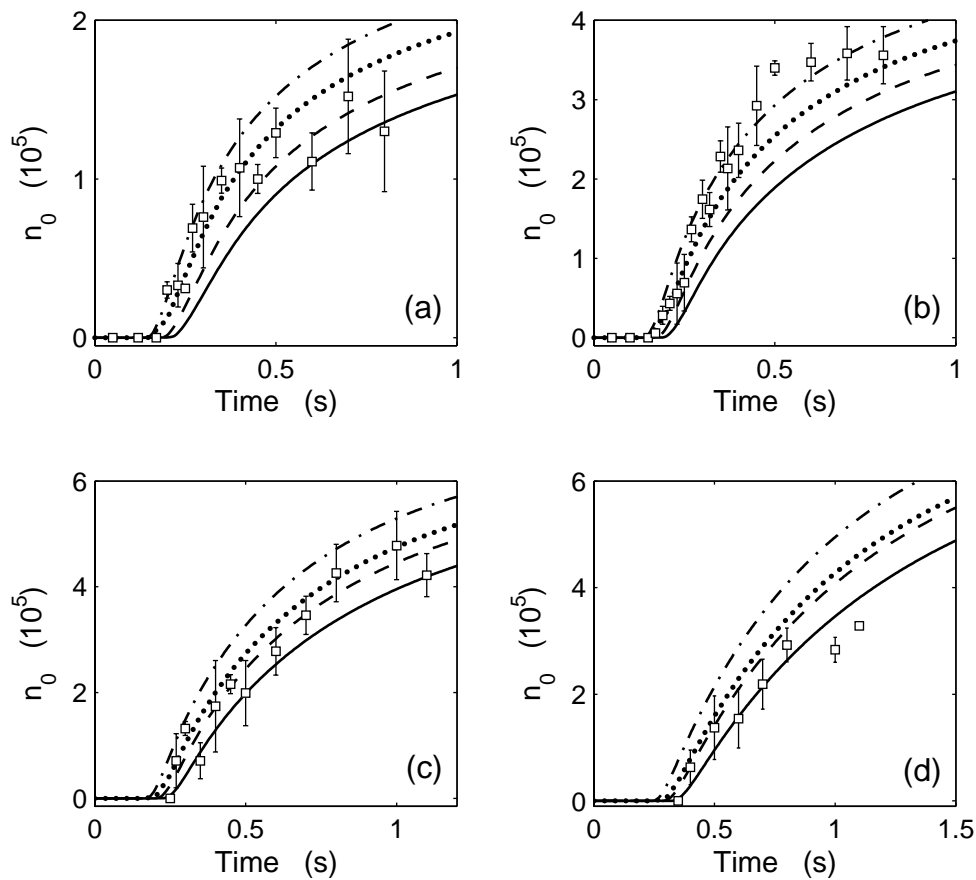}
\caption{\label{fig:expta} Comparison of theory and experiment for the growth 
of
a $^{87}$Rb Bose-Einstein condensate.  The squares with error bars indicate
the experimental data, while each line is the theoretical prediction based on 
a 
slightly different initial condition.  Each graph is for a different rf
frequency $\nu$.
Solid: $N_i = 4.2 \times 10^6$, $T_i = 640$ nK, $\muinit = -300 \hbar\bar{
\omega}$.
Dashed: $N_i = 4.4 \times 10^6$, $T_i = 640$ nK, $\muinit = -289 \hbar\bar{
\omega}$.
Dotted: $N_i = 4.2 \times 10^6$, $T_i = 610$ nK, $\muinit = -254 \hbar\bar{
\omega}$.
Dot-dash: $N_i = 4.4 \times 10^6$, $T_i = 610$ nK, $\muinit = -244 \hbar\bar{
\omega}$.
(a) $\nu = 2000$ kHz. 
(b) $\nu = 2015$ kHz. 
(c) $\nu = 2030$ kHz. 
(d) $\nu = 2045$ kHz. 
}
\end{figure}

\begin{figure}
\includegraphics{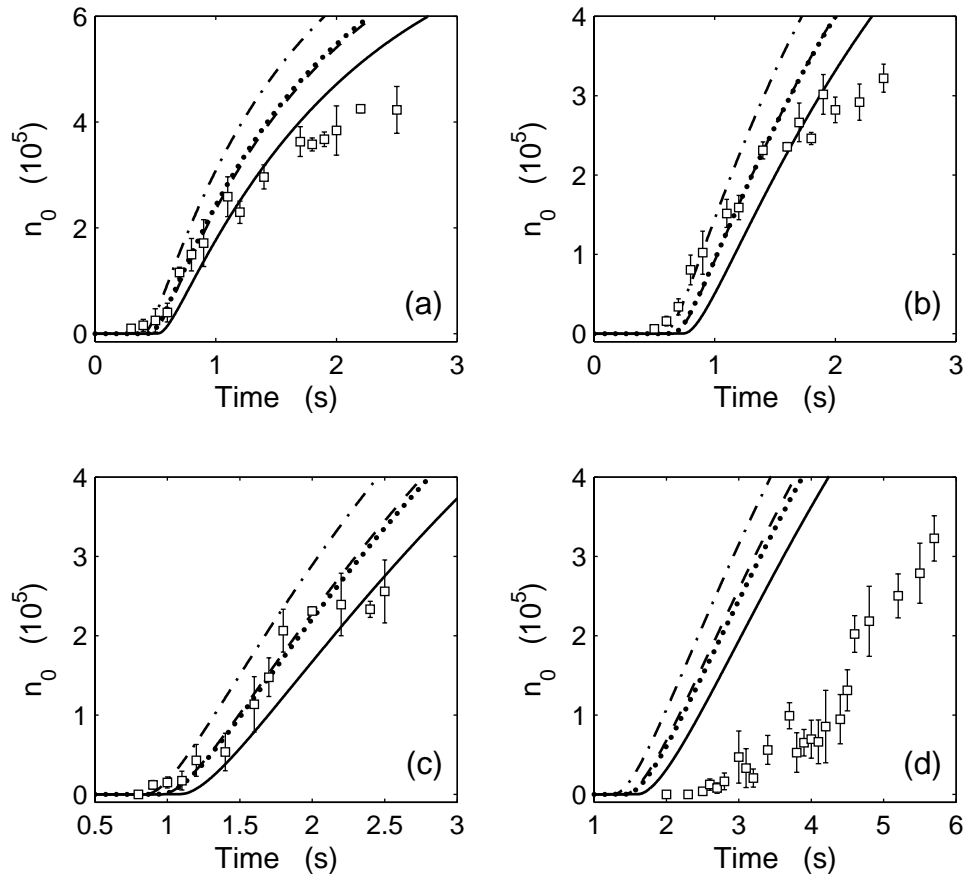}
\caption{\label{fig:exptb}Comparison of theory and experiment for the growth 
of
a $^{87}$Rb Bose-Einstein condensate.  The squares with error bars indicate
the experimental data, while each line is the theoretical prediction based on 
a 
slightly different initial condition.   Each graph is for a different rf
frequency $\nu$.
Solid: $N_i = 4.2 \times 10^6$, $T_i = 640$ nK, $\muinit = -300 \hbar\bar{
\omega}$.
Dashed: $N_i = 4.4 \times 10^6$, $ T_i = 640$ nK, $\muinit = -289 \hbar\bar{
\omega}$.
Dotted: $N_i = 4.2 \times 10^6$, $ T_i = 610$ nK, $\muinit = -254 \hbar\bar{
\omega}$.
Dot-dash: $N_i = 4.4 \times 10^6$, $ T_i = 610$ nK, $\muinit = -244 \hbar\bar{
\omega}$.
(a) $\nu = 2060$ kHz. 
(b) $\nu = 2070$ kHz. 
(c) $\nu = 2080$ kHz. 
(d) $\nu = 2090$ kHz.
}
\end{figure}

In this section we compare the results of our simulations as described above
with the experimental data provided by the Munich group.  In figures
\ref{fig:expta} and \ref{fig:exptb} we show the condensate growth data for all
experimental runs, along with four corresponding theoretical curves for each
run.  The solid lines are for the initial conditions reported by K\"{o}hl
\emph{et al}, and the others are within the statistical error with a chemical
potential slightly closer to zero (ie lower temperatures and larger number of
atoms).  Thus we do not present curves within the statistical error that begin
\emph{further} from the transition than the central value.  We do not include
background loss  in the simulations as the trap lifetime was more than 40
seconds \cite{michael_email}.


Considering the fact that there are \emph{no free parameters} in these
calculations, the fits of the theoretical curves to the experimental data are
impressive.  For the rf frequencies 2000--80 kHz the initiation times for
condensate growth are predicted extremely well, along with the initial rates of
condensate growth. It does seem for the slower growth curves with rf
frequencies 2060--80 kHz that the condensate occupation curve levels off
somewhat faster than the model predicts.  Unfortunately there is no 
experimental data at later times to determine whether the condensate continues
to grow. 

The cause of the flattening of these growth curves for these rf frequencies is
as yet undetermined.  It was originally suggested that there could be  a small
amount of heating present in the system that only becomes apparent in the
longer experiments.  However, careful analysis of the experimental data
\cite{michael_email} has ruled out this mechanism as an explanation for the
slow down of condensate growth.  A second possibility is that the three body
loss coefficient $K_3$ for this experimental set up may be different from that
reported in \cite{JILA_3body}; however, this is difficult to estimate.

The one instance where theory and experiment  differ strongly is for the 
slowest cooling experiment with an rf frequency of 2090 kHz, for which the
comparison is plotted in figure \ref{fig:exptb}(d).  The experimental data has
the peculiar feature that there is an apparent sudden increase in the growth
rate just over four seconds after the beginning of the experiment.  In
\cite{koehl} it is suggested that this jump is due to strong phase fluctuations
in the initial elongated condensate \cite{petrov}.   This feature, also known
as quasicondensation, was first suggested as a stage in the growth of a BEC  by
Kagan \etal \cite{kagan}. Recently phase fluctuations have been observed
experimentally in elongated condensates \cite{phase_fluc}.  

This behaviour can be explained physically as follows.  The initially strong
phase fluctuations can be thought of as the condensate having been seeded in
several parts, and thus bosonic stimulation occurs for each part
separately.  Therefore, the initial growth rate will be slower than would
otherwise be predicted theoretically.  As the phase coherence length grows,
however, suddenly a true condensate will form from the parts and there will be
a corresponding jump in the growth rate.

The lack of agreement between theory and the experimental data for the rf
frequency of 2090 kHz suggests that an effect not included in our model is
becoming important.
The experimentally observed initiation time is close to that predicted by the
theory, however the initial growth rate is much reduced.   Our model of
condensate growth makes the assumption that only a single condensate forms in
the system, and this is represented by a single quantum level.  Therefore, the
model currently 
cannot
represent slower initial growth 
due to the presence
of any quasicondensate. However, once the ``true'' condensate forms the 
observed  growth
rate does appear to be similar to that predicted by the simulations, but at a
later time. 

\section{\label{conclusion} Conclusions}

We have extended our model of condensate growth \cite{QKVII} to take into
account the sag of the atomic trapping potential due to gravity, and the 
effect this has on evaporative cooling at low temperature.  We have described
the effect this has on growth experiments, and carried out a comparison with
all available experimental data taken by the Munich group and presented in
\cite{koehl}.  We have found that despite the many approximations made, the
theoretical model is,  for the most part, in good agreement with the
experimental data.  It has been suggested for the one case in which there is a
discrepancy that this is due to the effects of quasicondensation---a
hypothesis that is not contradicted by our calculations.

\ack The authors would like to thank Michael K\"{o}hl for many useful
discussions and for the provision of the experimental data before publication. 
This work was supported by the Engineering and Physical Sciences Research
Council in the United Kingdom, and by the Royal Society of New Zealand under
the Marsden Fund Contract No. PVT-902.


\section*{References}
\begin{thebibliography}{99}

\bibitem{JILA} Anderson M,
Ensher JR, Matthews MR, Wieman CE and Cornell EA 1995 
{\it Science} \textbf{269} 198 

\bibitem{MIT} Davis KB,
Mewes M-O, Andrews MR, van~Druten NJ, Durfee DS, Kurn DM and Ketterle W 1995
{\it Phys.  Rev.  Lett.}  \textbf{75} 3969 

\bibitem{RICE} Bradley CC,
Sackett CA, Tollet JJ and Hulet R 1995
{\it Phys. Rev. Lett.} \textbf{75} 1687; 1997 {\it Phys. Rev. Lett.} \textbf{79} 1170 

\bibitem{stoof_long}
Stoof HTC 1999 {\it J. Low Temp. Phys.}
\textbf{114} 11  

\bibitem{QKT}
Gardiner CW and Zoller P 1997 \PR A \textbf{55} 2902;
Gardiner CW and Zoller P 1998 \PR A \textbf{58} 536;
Gardiner CW and Zoller P 2000 \PR A \textbf{61} 033601

\bibitem{MIT_growth} 
Miesner HJ, Stamper-Kurn DM, Andrews MR, Durfee DS, Inouye S and 
 Ketterle W 1998
{\it Science } \textbf{279} 1005 


\bibitem{BosGro1} 
Gardiner CW,  Ballagh RJ, Davis MJ and Zoller P 1997
{\it Phys. Rev. Lett.} \textbf{79} 1793

\bibitem{BosGro2} 
Gardiner CW, Lee MD, Ballagh RJ, Davis MJ and Zoller P 1998
{\it Phys. Rev. Lett.} \textbf{81} 5266 

\bibitem{QKVI}
Lee MD and Gardiner CW
{\it Phys. Rev.} A \textbf{62} 033606

\bibitem{QKVII}
Davis MJ, Gardiner CW, and Ballagh RJ
2000 \PR A \textbf{62} 063608

\bibitem{Bijlsma}
Bijlsma MJ, Zaremba E and Stoof HTC
2000 \PR A \textbf{62} 063609

\bibitem{koehl} K\"{o}hl M, Davis MJ, Gardiner CW, H\"{a}nsch T and Esslinger T 2001
\textit{Preprint} cond-mat/0106642

\bibitem{dalfovo}
Dalfovo F, Giorgini S, Pitaevskii LP and Stringari S 1999
\textit{Rev. Mod. Phys.} \textbf{71} 463 

\bibitem{ENS}
S{\"{o}}ding J, Gu{\'{e}}ry-Odelin D, Desbiolles P, Chevy F,
Inamori H and Dalibard J 1999
{\it Appl. Phys.} B \textbf{69} 257

\bibitem{JILA_3body}
Burt EA, Ghrist RW, Myatt CJ, Holland MJ, Cornell EA and Wieman CE 1997
{\it Phys. Rev. Lett.} \textbf{79} 337

\bibitem{michael_email} K\"{o}hl M 2001, 
\textit{Private communication} 

\bibitem{petrov} Petrov DS, Shlyapnikov GV and Walraven JTM
2001 \textit{Preprint} cond-mat/0104373

\bibitem{kagan} Kagan Yu, Svistunov BV and Shlyapnikov GV 1992 
\textit{Zh. Eksp. Teor. Fiz.} \textbf{101} 528 
(1992 \textit{Sov. Phys. JETP} \textbf{75} 387)

\bibitem{phase_fluc} Dettmer S, Hellweg D, Arlt J, Ertmer W, Sengstock K,
Petrov DS, Shlyapnikov GV, Kreutzmann H, Santos L and Lewenstein M 2001
\textit{Phys. Rev. Lett.} \textbf{87} 160406

\end{thebibliography}
\end{document}